# Large Universe Attribute-Based Encryption Scheme from Lattices

Shang-ping Wang[*]

School of science, Xi'an University of Technology
Xi'an, 710054, Shaanxi, CHINA
spwang99@hotmail.com

Fang Feng

School of science, Xi'an University of Technology
Xi'an, 710054, Shaanxi , CHINA
751770118@qq.com

(Lab of Cryptography and Network Security, Xi'an University of Technology, Xi'an, 710054, China)

**Abstract**. We propose a large universe attribute-based encryption (ABE ) scheme from lattices. It is inspired by Brent Waters's scheme which is a large universe attribute-based encryption using bilinear map. It is a very practical scheme but this scheme may not be security with the developing quantum computer. So we extend their good idea of large universe attribute-based encryption to lattices based cryptosystem. And our scheme is the first large universe ABE scheme from lattices. In a large universe ABE system any string can be used as attribute and attributes need not be determined at system setup. This is a desirable feature. And the master private key of our scheme is too short too a matrix. Moreover, our scheme is high efficient due to the ciphertext of our scheme is divided into three parts. Finally, under Learning with Errors (LWE) assumption, we prove our scheme is secure under the select attribute attack.

**Key words**：Large Universe; Lattices; Attribute-Based Encryption; Short key; Efficiency; Learning with Errors.

## 1. Introduction

The notion of Fuzzy Identity Based Encryption was first introduced by Sahai and Waters [1] at EUROCRYPT'05, which is the foundation of the emergence of attribute-based encryption (ABE) system. Attribute-Based Encryption has occupied a huge area of research in cryptography over the last few years. The emergence of Attribute-based encryption system makes the decryptor a group with the same properties instead of a single individual. Attribute-based encryption system makes the ciphertext can be shared by multi-user, so it improves access efficiency of the system. The most prominent advantage of the system is very applicable to the situations which not need to know the specific receiver, and the receiver who only satisfies the corresponding conditions can decrypt the message from sender. This makes the attribute-based encryption system be widely used.

The study of attribute based encryption schemes has already obtained many research results. Especially in recent years Waters et al. [2] made a great improvement of attribute-based encryption schemes in the aspect of practical, the most representative is proposed a scheme which supports large universe attribute encryption. In large universe attribute encryption the size of attribute universe can be exponentially large. This is a desirable feature compared the small universe attribute encryption which fixed attributes at system setup. But it is a pity all this practical scheme is based on bilinear pairing technology with the decisional bilinear Diffie-Hellman assumption or the variant decisional bilinear Diffie-Hellman assumption, such as literature [2], [3], [4]. But the main drawback of this type of scheme is generally inefficient, because the program is generally necessary to calculate multiple bilinear pairs. So people try to construct the attribute based cryptosystem using other techniques.

In recent years, many breakthroughs have been made in lattice to design cryptosystem, which mainly reflected in the construction of public key cryptosystems relies on lattice hardness problems. Cryptography scheme based on lattice theory has many advantages, such as high efficiency, simple structure which easy to understand. Moreover, the most important advantage is lattice-based schemes enjoying strong security guarantees. There is an intrinsic relation between the attribute based cryptosystem and fuzzy identity based cryptosystem, and there already have some literatures [5],[6],[7] designed several encryption schemes based on attribute from lattice. Although attribute based scheme on lattice has a little productive.

**Our motivation**: Combined with advantages and disadvantages of ABE from bilinear and the advantages and disadvantages of ABE from lattices, we hope to build a large universe ABE from lattices.

First of all, since Waters et al [2] propose a practical large universe ABE scheme recently which is a very practical scheme but the security of this scheme is threatened since the emergence of a quantum computer. So we hope to extend his good idea of large universe attribute-based encryption to lattices based cryptosystem.

Second, due to large universe ABE dose not determine the number of attributes in system setup, but we hope determine the master private key at once. This may avoid the flaws of modified the master private key after addition extra attributes later.

Finally, since in large universe ABE scheme, the number of attributes participates in the encryption may too large, so we hope message encryption can divide into different parts. Some parts have nothing to do with the number of attributes participate in the encryption, only one part related. Therefore, we can achieve efficient encryption

**Our Contributions**:

Our main result is the scheme of a large universe attribute-based encryption from lattices, and reduces its security from LWE.

First, our scheme is the first large universe ABE from lattices. We apply new idea access policy $(M, \rho)$ on LSSS which propose in Brent Waters's [2] scheme "Practical Constructions and New Proof Methods for Large Universe Attribute-Based Encryption" in 2013 to attribute-based encryption from lattices. And the main idea of this scheme is change the $LSSS$ access policy from one matrix form to a matrix and a $[l] \to Z_q$ map $\rho$ form. In previous $LSSS$ access policy, every row of the share-generating matrix $M$ is just map to an attribute. While in large universe ABE, map $\rho$ in access policy can map every row of the share-generating matrix $M$ to any attributes on universe. Through use the new access policy $(M, \rho)$ on lattices we get a large universe attribute-based encryption form lattices.

Second, our scheme has a short master key. The master private key of our scheme has only one matrix $\mathbf{T}_{A_0}$, which is a short base of public parameter $\mathbf{A}_0$. And the public key $\mathbf{E}_i = \mathbf{A}_0 \| \mathbf{A}_i + \mathbf{B}$ of every attribute is combined with two parts, the first part is the same public parameter $\mathbf{A}_0$, and the second part is $\mathbf{A}_i + \mathbf{B}$ which is determinant by every different attributes. So we hide our trapdoor $\mathbf{T}_{A_0}$ information just under the first part. Then the master private key is fixed no matter how many attributes have in our system. But the fault is that the size of ciphertext and the time of encryption will increase two times, since it is linear to the number of columns in $\mathbf{E}_i$.

Third, our scheme is efficient. In order to overcome the shortcoming above, we divide the ciphertext into three parts. And we not use public key $\mathbf{E}_i$ in encryption algorithm directly but divided $\mathbf{E}_i$ into two parts. So our ciphertext combine with three parts: the first part is message encryption, the second part is general encryption and the third part is attributes encryption, where the size of first and second is fixed that is irrelevant to the attributes participant encryption. The first part is only related to the message $msg$, the second general encryption part just related to the first part of public key $\mathbf{A}_0$, and the third part is encrypt by the second part $\mathbf{A}_i$ of public key. So will not repeat save and calculation the common part of first part $\mathbf{A}_0$ of public key. This allows the master private key into a matrix, but the size of the ciphertext and the encrypted time add not too much.

Finally, under LWE assumption, our scheme is security under the select attribute attack.

**Organization**：The rest of this paper is organized as follows. In section 2 we define our notion and recall some preliminaries. In section 3 we describe our construction. The Correctness and Security of our construction are given in section 4. Finally, we conclude in section 5.

## 2. Preliminaries

We first introduce the notion used throughout the paper, and then introduce the basic knowledge of lattice, some theorems on lattice. Finally we give the security assumptions used in our scheme.

## 2.1 Notation

Throughout this paper, we use uppercase boldface alphabet for matrices, as in $\mathbf{A}$; lowercase boldface characters for vectors, as in $\mathbf{e}$, and lowercase regular characters for scalars, as in $v, [l]$ be the set $\{1,2,\cdots,l\}$. Denoted $l_2$ norm by $\|\cdot\|$. For $\mathbf{X} \in R^{n \times m}$ and $\mathbf{Y} \in R^{n \times m'}$, $(\mathbf{X} \| \mathbf{Y}) \in R^{n \times (m+m')}$ denotes the concatenation of the column of $\mathbf{X}$ followed by the columns of $\mathbf{Y}$. Similarly, for $\mathbf{X} \in R^{n \times m}$ and $\mathbf{Y} \in R^{n' \times m}$, $(\mathbf{X}; \mathbf{Y}) \in R^{(n+n') \times m}$ is the concatenation of the rows of $\mathbf{X}$ followed by the rows of $\mathbf{Y}$.

## 2.2 Random Integer Lattices

**Definition1.** Let $\mathbf{B} = [\mathbf{b}_1 | \cdots | \mathbf{b}_m] \in R^{n \times m}$ be a $n \times m$ matrix whose columns are linearly independent vectors $\mathbf{b}_1, \cdots, \mathbf{b}_m \in R^n$. The m-dimensional $\Lambda$ generated by $\mathbf{B}$ is infinite periodic set,

$$\Lambda = L(\mathbf{B}) = \left\{ y \in R^m \ s.t.\ \exists s = (s_1, \cdots, s_m) \in Z^m, \mathbf{y} = \mathbf{Bs} = \sum_{i=1}^{m} s_i \mathbf{b}_i \right\}$$

Here, we are interested in integer lattices, i.e. infinite periodic subsets of $Z^m$.

**Definition2.** For $q$ and $\mathbf{A} \in Z_q^{n \times m}$ and $\mathbf{u} \in Z_q^n$, define:

$$\Lambda_q^{\perp}(\mathbf{A}) = \{\mathbf{e} \in Z^m \ s.t.\ \mathbf{Ae} = 0 \ (\bmod q)\}$$

$$\Lambda_q^{\mathbf{u}}(\mathbf{A}) = \{\mathbf{e} \in Z^m \ s.t.\ \mathbf{Ae} = \mathbf{u} \ (\bmod q)\}$$

## 2.3 Trapdoors for Lattices: The algorithm TrapGen

Ajtai [8] showed how to sample an essentially uniform matrix $\mathbf{A} \in Z_q^{n \times m}$ with an associated full-rank set $\mathbf{T_A} \subset \Lambda^{\perp}(\mathbf{A})$ of low-norm vectors. We will use an improved version of Ajtai's basis sampling algorithm due to Alwen and Peikert [9]:

**Definition3** ([9]). Let $n = n(\lambda), q = q(\lambda), m = m(\lambda)$ be positive integers with $q \geq 2$ and $m \geq 5n \log q$. There exists a probabilistic polynomial-time algorithm TrapGen that outputs a pair $\mathbf{A} \in Z_q^{n \times m}, \mathbf{T_A} \in Z_q^{n \times m}$ such that $\mathbf{A}$ is statistically close to uniform and $\mathbf{T_A}$ is a basis for $\Lambda^{\perp}(\mathbf{A})$ with length $L = \|\widetilde{\mathbf{T}}_\mathbf{A}\| \leq m \cdot \omega(\sqrt{\log m})$ with all but $n^{-\omega(1)}$ probability.

## 2.4 Discrete Gaussians

**Definition4.** Let $m \in Z_{>0}$ be a positive integer and $\Lambda \in R^m$ an m-dimensional lattice. For any vector $\mathbf{c} \in R^m$ and any positive parameter $\sigma \in R_{>o}$, we define:

$\rho_{\sigma,\mathbf{c}}(\mathbf{x}) = \exp\left(-\pi \frac{\|\mathbf{x}-\mathbf{c}\|^2}{\sigma^2}\right)$ : A Gaussian-shaped function on $R^m$ with center $\mathbf{c}$ and parameter $\sigma$,

$\rho_{\sigma,\mathbf{c}}(\Lambda) = \sum_{\mathbf{x}\in\Lambda} \rho_{\sigma,\mathbf{c}}(\mathbf{x})$: The (always converging) discrete integral of $\rho_{\sigma,\mathbf{c}}$ over the lattice $\Lambda$,

$D_{\Lambda,\sigma,\mathbf{c}}$: The discrete Gaussian distribution over $\Lambda$ with center $\mathbf{c}$ and parameter $\sigma$,

$$\forall \mathbf{y} \in \Lambda, \ D_{\Lambda,\sigma,c}(\mathbf{y}) = \frac{\rho_{\sigma,c}(\mathbf{y})}{\rho_{\sigma,c}(\Lambda)}$$

For notional convenience, $\rho_{\sigma,0}$ and $D_{\Lambda,\sigma,0}$ are abbreviated as $\rho_\sigma$ and $D_{\Lambda,\sigma}$.

## 2.5 Gaussians Sampling algorithms over Lattices

The following Gaussians Sampling algorithms over lattices are used in basic construction and security proof.

### 2.5.1 **SampleLeft algorithms**

**SampleLeft algorithms** [10]: let $q > 2$, $m > n$. There is an algorithm SampleLeft$(\mathbf{A},\mathbf{B},T_\mathbf{A},s,\mathbf{u})$ takes a full rank matrix $\mathbf{A} \in Z_q^{n\times m}$, a matrix $\mathbf{B} \in Z_q^{n\times m_1}$, a basis $T_\mathbf{A}$ of $\Lambda_q^u(\mathbf{A})$, a Gaussian parameter $s > \|\tilde{T}_\mathbf{A}\| \cdot \omega\left(\sqrt{\log(m+m_1)}\right)$ and a vector $\mathbf{u}$ as input, output a vector $\mathbf{e} \in Z_q^{m+m_1}$ distributed statistically close to $D_{\Lambda_q^\mathbf{u}(\mathbf{F}),s}$, where $\mathbf{F} := (\mathbf{A} \| \mathbf{B})$.

### 2.5.2 Preimage Sampling

We will use following algorithm from [11]. Let $q \geq 2$, $m \geq 2n\log q$.

Algorithm SamplePre$(\mathbf{A},\mathbf{T}_\mathbf{A},\mathbf{u},\sigma)$[11]: On input a matrix $\mathbf{A} \in Z_q^{n\times m}$ with 'short' trapdoor basis $\mathbf{T}_\mathbf{A}$ for $\Lambda_q^\mathrm{T}(\mathbf{A})$, a target image $\mathbf{u} \in Z_q^n$ and a Gaussian parameter $\sigma \geq \|\tilde{\mathbf{T}}_\mathbf{A}\| \cdot \omega(\sqrt{\log m})$, outputs a sample $\mathbf{e} \in Z^m$ from a distribution that is within negligible statistical distance $D_{\Lambda_q^\mathbf{u}(\mathbf{A}),\sigma}$.

### 2.5.3 Lattice basis extension algorithm

*ExtBasis* [12]: There is a deterministic polynomial-time algorithm *ExtBasis*$(\mathbf{S}, \mathbf{A}' = \mathbf{A} \| \overline{\mathbf{A}})$ with following properties: given an arbitrary matrix $\mathbf{A} \in Z_q^{n\times m}$ whose columns generate group $Z_q^n$,

an arbitrary basis $\mathbf{S} \in Z^{m \times m}$ of $\Lambda^{\perp}(\mathbf{A})$ and let $\overline{\mathbf{A}} \in Z_q^{n \times \overline{m}}$ be arbitrary. $ExtBasis(\mathbf{S}, \mathbf{A}' = \mathbf{A} \| \overline{\mathbf{A}})$ Outputs a basis $\mathbf{S}'$ of $\Lambda^{\perp}(\mathbf{A}') \in Z^{m+\overline{m}}$ such that $\|\widetilde{\mathbf{S}}'\| = \|\widetilde{\mathbf{S}}\|$.

Moreover, the statement holds even if the columns of $\mathbf{A}'$ are permuted arbitrarily ( $\mathbf{A} \| \overline{\mathbf{A}}$ denote the concatenation of matrix $\mathbf{A}$ and $\overline{\mathbf{A}}$ )

## 2.6 Access Structures and Linear Secret-Sharing Schemes

In this section, we present the formal definition of access structures and linear secret-sharing scheme, adapted to match our setting.

**Definition 5** (Access Structures) [13]. Let $U$ be the attributes universe. An access structure on $U$ is a collection $A$ of non-empty sets of attributes, i.e. $A \subseteq 2^U \setminus \{\ \}$. The sets in $A$ are called the authorized sets and the sets not in $A$ are called the unauthorized sets.

Additionally, an access structure is called monotone if $\forall B, C \in U$: if $B \in A$ and $B \subseteq C$, then $C \in A$.

In our constructions, we only consider monotone access structures, which mean that a user acquires more attributes; he will not lose his possible decryption privileges.

**Definition 6** (Linear Secret-Sharing Schemes (LSSS)) [13]. Let $q$ be a prime and $U$ the attributes universe. A secret-sharing scheme $\Pi$ with domain of secrets $Z_q$ realizing access structures on $U$ is linear over $Z_q$ if

1. The shares of a secret $s \in Z_q$ for each attribute form a vector over $Z_q$.

2. For each access structure $A$ on $U$, there exists a matrix $M \in Z_q^{l \times n}$, called the share-generating matrix, and a function $\rho$, that labels the rows of $M$ with attributes from $U$, i.e. $\rho : [l] \to U$, which satisfy the following: During the generation of shares, we consider the column vector $\mathbf{v} = (s, r_2, \cdots r_n)^T$, where $r_2, \cdots r_n \xleftarrow{\$} Z_q$. Then the vector of $l$ shares of the secret $s$ according to $\Pi$ is equal to $M\mathbf{v} \in Z_q^l$. The share $(M\mathbf{v})_j$ where $j \in [l]$ "belong" to attribute $\rho(j)$.

We will be referring to the pair $(M, \rho)$ as the policy of the access structure $A$.

According to [13], each secret-sharing scheme should satisfy the reconstruction requirement for the authorized sets and any unauthorized set cannot reveal any partial information about the secret. In our setting, let $S$ denote an authorized set for the access structure $A$ encoded by the policy $(M,\rho)$. Then let $I$ be the set of rows whose labels are in $S$, i.e. $I = \{i \mid i \in [l] \land \rho[i] \in S\}$. The reconstruction requirement asserts that the vector $(1,0,\cdots,0)$ is in the span of rows of $M$ indexed by $I$. This means that there exist constants $\{w_i\}_{i \in I}$ such that for any valid shares $\{\lambda_i = (M\mathbf{v})_i\}_{i \in I}$ of a secret $s$ according to $\Pi$, it is true that $\sum_{i \in I} w_i \lambda_i = s$. Additionally, it has been proved in [13] that the constants $\{w_i\}_{i \in I}$ can be found in time polynomial in the size of the share-generating matrix $M$

## 2.7 Hardness Assumption

The $LWE$ (learning with errors) problem was first defined by Oded Regev [14] in 2005, and has since been extensively studied and used. We use the decisional version of the $LWE$ problem.

**Definition 7.** Consider a prime $q$, a positive integer $n$, and a distribution $\chi \in Z_q$, all public. An $(Z_q, n, \chi) - LWE$ problem instance consists of access to an unspecified challenge oracle $O$, being, either, a noisy pseudo-random sampler $O_\mathbf{x}$ carrying some constant random secret key $\mathbf{x} \in Z_q^n$, or, a truly sampler $O_\$$, whose behaviors are respectively as follows:

$O_\mathbf{x}$: Output noisy pseudo-random samples of the form $(w_i, v_i) = (w_i, w_i^T \mathbf{x} + \chi_i) \in Z_q^n \times Z_q$, where, $\mathbf{x} \in Z_q^n$ is a uniformly distributed persistent secret key that is invariant across invocations. $\chi_i$ is a freshly generated ephemeral additive noise component with distribute $\chi$, and $w_i \in Z_q^n$ is a fresh uniformly distributed vector revealed as part of the output.

$O_\$$: Output truly random samples $(w_i, v_i) \in Z_q^n \times Z_q$, drawn in independently uniformly at random in the entire domain $Z_q^n \times Z_q$.

The $(Z_q, n, \chi) - LWE$ problem statement, or $LWE$ for short, allows an unspecified number of queries to be make to the challenge oracle $O$, with no stated or prior bound. We say that an algorithm $A$ decides the $(Z_q, n, \chi) - LWE$ problem if $\left|\Pr[A^{O_\mathbf{x}} = 1] - \Pr[A^{O_\$}] = 1\right|$ is non-negligible for a random $\mathbf{x} \in Z_q^n$.

It has been shown in [14] that there is a $poly(n,q) - time$ reduction from

Search $(Z_q, n, \chi) - LWE$ to decision $(Z_q, n, \chi) - LWE$.

The confidence in the hardness of the $LWE$ problem stems in part of a result of Regev[14] which shows that the for certain noise distributions $\chi$, the $LWE$ problem is as hard as the worst-case $SIVP$ and $GapSVP$ under a quantum reduction. A classical reduction with related parameters was later obtained by Peikert[5].

**Proposition** ([14], Theorem1). Consider a real parameter $\alpha = \alpha(n) \in (0,1)$ and a prime $q = q(n) > 2\sqrt{n}/\alpha$. Denote by $T = R/Z$ the group of reals $[0,1)$ with addition modulo 1. Denote by $\Psi_a$ the distribution over T of a normal variable with mean 0 and standard deviation $2\sqrt{n}/\alpha$ then reduced modulo 1. Denote by $\lfloor x \rceil = \lfloor x + \frac{1}{2} \rfloor$ the nearest integer to the real $x \in R$.

Denote by $\overline{\Psi}_\alpha$ the discrete distribution over $Z_q$ of the random variable $\lfloor qX \rceil \bmod q$ where the random variable $X \in T$ has distribution $\Psi_a$.

Then if there exists an efficient, possibly quantum, algorithm for deciding the $(Z_q, n, \overline{\Psi}_\alpha) - LWE$ problem, there exists a quantum $q \cdot poly(n)$-time algorithm for approximating the $SIVP$ and $GapSVP$ problem, to within $\tilde{O}(n/\alpha)$ factors in the $l_2$ norm, in the worst case.

Since the best known algorithms for $2^k$-approximations of $gapSVP$ and $SIVP$ run in time $2^{\tilde{O}(n/k)}$[7], it follows from the above that the $LWE$ problem with the noise ratio $\alpha = 2^{-n^\varepsilon}$ is likely hard for some constant $\varepsilon < 1$.

Next two lemmas will need to show that decryption works correctly.

**Lemma 9** [15]. For any n-dimension lattice $\Lambda$, vector $\mathbf{c} \in R^n$, and real $\varepsilon \in (0,1), s > \eta_\varepsilon(\Lambda)$, we have

$$\Pr_{\mathbf{x} \sim D_{\Lambda,s,\mathbf{c}}} \left[ \|\mathbf{x} - \mathbf{c}\| > s\sqrt{n} \right] \leq \frac{1-\varepsilon}{1+\varepsilon} \cdot 2^{-n}$$

The lemma sates that for large enough $s$, almost the elements chosen from $D_{\Lambda,s,\mathbf{c}}$ are close to $\mathbf{c}$.

**Lemma 10**[16]. Let $\mathbf{e}$ be some vector in $Z_m$ and let $\mathbf{y} \leftarrow \overline{\Psi}_\alpha^m$. Then the quantity $|\mathbf{e}^T \mathbf{y}|$ treated as an integer in $[0, q-1]$ satisfies

$$|e^T \mathbf{y}| \leq \|\mathbf{e}\| q \alpha \omega(\sqrt{\log m}) + \|\mathbf{e}\| \sqrt{m}/2$$

With all but negligible probability in *m*. In particularly, if $x \leftarrow \overline{\Psi}_\alpha$ is treated as an integer in $[0, q-1]$ then $|x| \leq q \alpha \omega(\sqrt{\log m}) + 1/2$ with all but negligible probability in *m*.

## 3. A Large Universe Attribute-Based Encryption Scheme from Lattices

Our scheme is inspired by large universe ABE on bilinear and ciphertext policy ABE from lattices proposed in [2] and [7] respectively. And combine them we achieve our scheme large universe ABE from lattices. The concrete construction consists of the following four PPT algorithms:

*Setup.*$(1^\lambda)$. Takes the security parameter $\lambda$ and a description of attributes universe $U = Z_q$ as input, and outputs the public parameters *pp* and the master private key *msk*. Where prime $q = \Theta(2^\lambda)$.

1. Use algorithm $\text{TrapGen}(1^\lambda)$ to select a uniformly random $n \times m$ matrix $\mathbf{A}_0 \in Z_q^{n \times m}$ with full-rank *m*-vector base $T_{\mathbf{A}_0} \in Z_q^{m \times m}$ such that $T_{\mathbf{A}_0} \in \Lambda_q^\perp(\mathbf{A}_0)$ and $\|\tilde{T}_{A_0}\| \leq m \cdot \omega(\sqrt{\log m})$;

2. Select a matrix $B \in Z_q^{n \times m}$ randomly;

3. For each attributes $i \in U$, select matrix $\mathbf{A}_i \in Z_q^{n \times m}$ randomly;

4. Select a uniform random *l*-vector $\mathbf{s}^T = (s_1, s_2, \cdots s_l) \in Z_q^l$;

5. Return the public parameter $pp = (\mathbf{A}_0, \mathbf{B}, \{\mathbf{A}_i\}_{i \in U}, \mathbf{s})$ and the master private key $msk = \mathbf{T}_{A_0}$.

*KeyGen*$(pp, msk, (M, \rho))$. On input the public parameter *pp*, the master private key $msk = \mathbf{T}_{A_0}$ and an access policy $(M, \rho)$, where $M \in Z_q^{l \times n}, \rho : [l] \to Z_q$, output the private key *sk*.

1. Select vector $\mathbf{y}_1, \mathbf{y}_2, \cdots \mathbf{y}_l \in Z_q^n$ such that $\mathbf{y}_1 = (s_1, y_{12}, y_{13}, \cdots y_{1n})^T; \mathbf{y}_2 = (s_2, y_{22}, \cdots, y_{2n})^T \cdots; \mathbf{y}_l = (s_l, y_{l2}, \cdots, y_{ln})^T$, where $s_1, s_2, \cdots s_l \in Z_q$ is the corresponding components of $\mathbf{s}^T = (s_1, s_2, \cdots s_l) \in Z_q^l$, and $y_{i,j} \in Z_q$ is choose randomly for $i = 1, \cdots l; j = 1, 2, \cdots, n$;

2. Let $M\mathbf{y}_i = \left(\lambda_1^{(i)}, \lambda_2^{(i)}, \cdots \lambda_l^{(i)}\right)^T, i = 1, \cdots l$.

3. Use SampleLeft algorithm get the private key of attributes $\rho(i)$, for $i = 1, \cdots l$. let

$$\boldsymbol{\lambda}_i^{(j)} = \left(\underbrace{0, \cdots, 0}_{j-1 \uparrow 0}, \lambda_i^{(j)}, \underbrace{0, \cdots, 0}_{n-j \uparrow 0}\right)^T \quad \text{and} \quad \mathbf{E}_{\rho(i)} = \mathbf{A}_0 \| \mathbf{A}_{\rho(i)} + \mathbf{B}, \quad \text{then call the}$$

algorithm $\text{SamplLefte}\left(\mathbf{A}_0, \mathbf{A}_{\rho(i)} + \mathbf{B}, \sigma, \boldsymbol{\lambda}_i^{(j)}\right)$ get $\mathbf{e}_{\rho(i)}^{(j)} \in Z_q^{2m \times 1}$, i.e.

$$\mathbf{E}_{\rho(i)} \mathbf{e}_{\rho(i)}^{(j)} = \left(\underbrace{0, \cdots, 0}_{j-1 \uparrow 0}, \lambda_i^{(j)}, \underbrace{0, \cdots, 0}_{n-j \uparrow 0}\right)^T \in Z_q^n; \text{where } \boldsymbol{\lambda}_i^{(j)} = \left(\underbrace{0, \cdots, 0}_{j-1 \uparrow 0}, \lambda_i^{(j)}, \underbrace{0, \cdots, 0}_{n-j \uparrow 0}\right), \text{ namely}$$

$$\mathbf{E}_{\rho(1)} \mathbf{e}_{\rho(1)}^{(1)} = \left(\lambda_1^{(1)}, 0, \cdots 0\right)^T; \mathbf{E}_{\rho(1)} \mathbf{e}_{\rho(1)}^{(2)} = \left(0, \lambda_1^{(2)}, \cdots 0\right)^T \cdots; \mathbf{E}_{\rho(1)} \mathbf{e}_{\rho(1)}^{(l)} = \left(0, 0, \cdots \lambda_1^{(l)}, \cdots, 0\right)^T$$

$$\mathbf{E}_{\rho(2)} \mathbf{e}_{\rho(2)}^{(1)} = \left(\lambda_2^{(1)}, 0, \cdots 0\right)^T; \mathbf{E}_{\rho(2)} \mathbf{e}_{\rho(2)}^{(2)} = \left(0, \lambda_2^{(2)}, \cdots 0\right)^T; \cdots; \mathbf{E}_{\rho(2)} \mathbf{e}_{\rho(2)}^{(l)} = \left(0, 0, \cdots \lambda_2^{(l)}, \cdots, 0\right)^T$$

$$\ddots$$

$$\mathbf{E}_{\rho(l)} \mathbf{e}_{\rho(l)}^{(1)} = \left(\lambda_l^{(1)}, 0, \cdots, 0\right)^T; \mathbf{E}_{\rho(l)} \mathbf{e}_{\rho(l)}^{(2)} = \left(0, \lambda_l^{(2)}, \cdots, 0\right)^T; \cdots; \mathbf{E}_{\rho(l)} \mathbf{e}_{\rho(l)}^{(l)} = \left(0, 0 \cdots \lambda_l^{(l)}, 0 \cdots, 0\right)^T$$

Where $\sigma \geq m \cdot \omega(\log m) \geq \|\tilde{T}_{\mathbf{A}_0}\| \cdot \omega\left(\sqrt{\log m}\right)$, then according to the Algorithm $\text{SamplLefte}\left(\mathbf{A}_0, \mathbf{A}_{\rho(i)} + \mathbf{B}, \sigma, \boldsymbol{\lambda}_i^{(j)}\right)$ we know $\mathbf{e}_{\rho(i)}^{(j)}$ from a distribution which close to the distribution $D_{\Lambda_q^{\lambda_i^{(j)}}(\mathbf{A}_i), \sigma}$. So we know $\|\mathbf{e}_i^{(j)}\| \leq \sigma \sqrt{m}$ with overwhelming probability.

4. Obtain the private key $sk_{\rho(i)} = \left(\mathbf{e}_{\rho(1)}^{(i)}, \mathbf{e}_{\rho(2)}^{(i)}, \cdots, \mathbf{e}_{\rho(l)}^{(i)}\right)$ of attributes $\rho(i)$, and set the private key $sk_{\rho(i)} = \left(\mathbf{e}_{\rho(1)}^{(i)}, \mathbf{e}_{\rho(2)}^{(i)}, \cdots, \mathbf{e}_{\rho(l)}^{(i)}\right)$ to the entity who has the attributes $\rho(i)$ under access policy $(M, \rho)$ through secure channel.

Note: Under a access policy $(M, \rho)$, the decrypt key of an attributes list $S$ is

$$sk = \left((M, \rho), \left(sk_{\rho(i)}\right)_{i \in [S]}\right).$$

$Encrypt(pp, S, msg)$: On input the public parameter $pp$, an attributes set $S \subseteq U = Z_p$ and a plaintext $msg \in \{0, 1\}$, output ciphertext $ct$.

1. Choose a uniformly random $n$-vector $\mathbf{x} \in Z_q^n$.

2. Constructs a $n$ dimension vector $\mathbf{f}$, just expands $l$ dimension vector $\mathbf{s}^T = (s_1, s_2, \cdots s_l) \in Z_q^l$ in the public parameter $pp$ to $n$ dimension. Set $f_i = s_i, i \leq l; f_i = 0, l < i \leq n$;

3. Select a low-norm Gaussian noise scalar $\chi_0 \in Z_q$ according to some parametric distribution $\overline{\Psi}_\alpha$, and select a low-norm Gaussian noise vector $\boldsymbol{\chi}'$, $\boldsymbol{\chi}_i \in Z_q^{2m}$ from $(\overline{\Psi}_\alpha)^m$ for every $i \in S$; And compute:

$$C_0 = \mathbf{x}^T \mathbf{f} + \chi_0 + \left\lfloor \frac{q}{2} \right\rfloor msg \in Z_q$$

$$C' = \mathbf{x}^T \mathbf{A}_0 + \boldsymbol{\chi}' \in Z_q^{2m}$$

$$C_i = \mathbf{x}^T (\mathbf{A}_i + B) + \boldsymbol{\chi}_i^T \in Z_q^{2m}$$

Then output the ciphertext $ct = (S, C_0, C', \{C_i\}_{i \in S})$.

$Decrypt(pp, sk, ct)$: On input the public parameter $pp$, the private key $sk$ about an attributes set and a ciphertext $ct$. Output the message $msg$.

First the decryption algorithm calculates the set of rows in $M$ that provide a share to attributes in $S$, i.e. $I = \{i : \rho(i) \in S\}$. Then it computes the constants $\{\omega_i \in Z_q\}_{i \in I}$ such that $\sum_{i \in I} \omega_i M_i = (1, 0, \cdots 0)$, where $M_i$ is $i$-th row of the matrix $M$. These constants exist if the set $S$ is an authorized set of the policy (see [13]).

Then it calculates $r = C_0 - \sum_{j=1}^{l} \sum_{i \in I} \omega_i (C'; C_i) \mathbf{e}_{\rho(i)}^{(j)}$, If $|r| < \frac{q}{4}$, output 0, else output 1.

## 4. Correctness and Security

### 4.1 Correctness

In this subsection, we show that our construction is correct with some appropriate parameters setting.

**Theorem 11** if we select the appropriate parameter, the Attribute-Based Encryption Scheme from Lattices in the previous section will be decryption successfully with overwhelming probability.

**Proof** If the decryptor can calculate $\sum_{i \in I} \omega_i M_i = (1, 0, \cdots 0)$, and know the private key $sk = \left( (M, \rho), (sk_{\rho(i)})_{i \in [S]} \right)$, then decrypted by the calculation:

$$(C'; C_i) = \begin{pmatrix} C' \\ C_i \end{pmatrix} = \begin{pmatrix} \mathbf{x}^T \mathbf{A}_0 + \chi'^T \\ \mathbf{x}^T (\mathbf{A}_i + B) + \chi_i^T \end{pmatrix}$$

$$= \begin{pmatrix} \mathbf{x}^T \mathbf{A}_0 \\ \mathbf{x}^T (\mathbf{A}_i + B) \end{pmatrix} + \begin{pmatrix} \chi'^T \\ \chi_i^T \end{pmatrix}$$

$$= \mathbf{x}^T (\mathbf{A}_0 \| (\mathbf{A}_i + B)) + \begin{pmatrix} \chi'^T \\ \chi_i^T \end{pmatrix}$$

And set $\chi = (\chi'^T; \chi_i^T)$

$$r = C_0 - \sum_{j=1}^{l} \sum_{i \in I} \omega_i (C'; C_i) \mathbf{e}_{\rho(i)}^{(j)}$$

$$= C_0 - \sum_{j=1}^{l} \sum_{i \in I} \omega_i \left( \mathbf{x}^T (\mathbf{A}_0 \| (\mathbf{A}_i + B)) + \chi \right) \mathbf{e}_{\rho(i)}^{(j)}$$

$$= C_0 - \sum_{j=1}^{l} \sum_{i \in I} \omega_i \left( \mathbf{x}^T (\mathbf{A}_0 \| (\mathbf{A}_i + B)) \right) \mathbf{e}_{\rho(i)}^{(j)} - \sum_{j=1}^{l} \sum_{i \in I} \omega_i (\chi) \mathbf{e}_{\rho(i)}^{(j)}$$

$$= C_0 - \mathbf{x}^T \sum_{j=1}^{l} \sum_{i \in I} \omega_i \left( (\mathbf{A}_0 \| (\mathbf{A}_i + B)) \right) \mathbf{e}_{\rho(i)}^{(j)} - \sum_{j=1}^{l} \sum_{i \in I} \omega_i (\chi) \mathbf{e}_{\rho(i)}^{(j)}$$

$$= C_0 - \mathbf{x}^T \sum_{j=1}^{l} \sum_{i \in I} \omega_i \mathbf{E}_i \mathbf{e}_{\rho(i)}^{(j)} - \sum_{j=1}^{l} \sum_{i \in I} \omega_i (\chi) \mathbf{e}_{\rho(i)}^{(j)} \quad (1)$$

$$= C_0 - \mathbf{x}^T \sum_{j=1}^{l} \sum_{i \in I} \omega_i \left( \underbrace{0, \cdots, 0}_{j-1 \uparrow 0}, \lambda_i^{(j)}, \underbrace{0, \cdots, 0}_{n-j \uparrow 0} \right)^T - \sum_{j=1}^{l} \sum_{i \in I} \omega_i (\chi) \mathbf{e}_{\rho(i)}^{(j)} \quad (2)$$

$$= C_0 - \mathbf{x}^T \sum_{j=1}^{l} (0, 0 \cdots s_j, \cdots 0)^T - \sum_{j=1}^{l} \sum_{i \in I} \omega_i (\chi) \mathbf{e}_{\rho(i)}^{(j)} \quad (3)$$

$$= C_0 - \mathbf{x}^T \mathbf{f} - \sum_{j \in I} \left( \sum_{i=1}^{l} \omega_i (\chi) \mathbf{e}_i^{(j)} \right)$$

$$= \left( \mathbf{x}^T \mathbf{f} + \chi_0 + \left\lfloor \frac{q}{2} \right\rfloor msg \right) - \mathbf{x}^T \mathbf{f} - \sum_{j \in I} \left( \sum_{i=1}^{l} \omega_i (\chi) \mathbf{e}_i^{(j)} \right)$$

$$= \left\lfloor \frac{q}{2} \right\rfloor msg + \chi_0 - \sum_{j \in I} \left( \sum_{i=1}^{l} \omega_i (\chi) \mathbf{e}_i^{(j)} \right)$$

Where (1)、(2)、(3) we use the fact $\mathbf{E}_{\rho(i)} \mathbf{e}_{\rho(i)}^{(j)} = \left( \underbrace{0, \cdots, 0}_{j-1 \uparrow 0}, \lambda_i^{(j)}, \underbrace{0, \cdots, 0}_{n-j \uparrow 0} \right)^T \in Z_q^n$, $\sum_{i \in I} \omega_i M_i = (1, 0, \cdots 0)$, in fact because

$M \mathbf{y}_i = \left( \lambda_1^{(i)}, \lambda_2^{(i)}, \cdots \lambda_l^{(i)} \right)^T$, $\mathbf{y}_i = (s_i, y_{i2}, \cdots, y_{in})^T$ then

$$\sum_{i\in I} \omega_i \left( \underbrace{0,\cdots,0}_{j-1\uparrow 0}, \lambda_i^{(j)}, \underbrace{0,\cdots,0}_{n-j\uparrow 0} \right)^{\mathrm{T}}$$

$$= \sum_{i\in I} \omega_i \left( \underbrace{0,\cdots,0}_{j-1\uparrow 0}, M_i \mathbf{y}_j, \underbrace{0,\cdots,0}_{n-j\uparrow 0} \right)^{\mathrm{T}}$$

$$= \left( \underbrace{0,\cdots,0}_{j-1\uparrow 0}, \sum_{i\in I} \omega_i M_i \mathbf{y}_j, \underbrace{0,\cdots,0}_{n-j\uparrow 0} \right)^{\mathrm{T}}$$

$$= \left( \underbrace{0,\cdots,0}_{j-1\uparrow 0}, s_j, \underbrace{0,\cdots,0}_{n-j\uparrow 0} \right)^{\mathrm{T}}$$

## 4.2 Security

In this subsection, we prove our scheme's security in the standard model under above parameters setting.

**Theorem 12** If there is a probabilistic polynomial-time algorithm $\mathcal{A}$ with advantage $\varepsilon > 0$ in a selective-security attribute list attack against the above scheme, Then there exists a probabilistic polynomial-time algorithm $\mathcal{B}$ that decides the $(Z_q, n, \overline{\psi}_\alpha) - LWE$ problem with advantage $\varepsilon$, where $\alpha = O(poly(n))$.

**Proof.** In the $LWE$ problem, the decision algorithm is given access to a sampling oracle, $O$, which is either a pseudo-random $O_\mathbf{x}$ with embedded secret $\mathbf{x} \in Z_q^n$, or a truly random sampler $O_\$$ (see **definition** 7), Our decide algorithm $\mathcal{B}$ will simulate an attack environment for, and exploit the prowess of $\mathcal{A}$, to decide which oracle it is given. The reduction proceeds as follows.

**Target.** $\mathcal{A}$ announces a target attribute set $S^* = \{\text{Attrib}_1, \text{Attrib}_2, \cdots \text{Attrib}_t\} \subseteq U$, on which it wishes to be challenged.

**Setup.** Challenger $\mathcal{B}$ get target attribute set from adversary $\mathcal{A}$.

1. $\mathcal{B}$ draws $(\mathbf{s}, v_s) \in Z_q^n \times Z_q$, $(\mathbf{A}_0, \mathbf{v}_0) \in Z_q^{n\times m} \times Z_q^n$ from $O$. And compute $(\mathbf{B}, \mathbf{T_B}) \leftarrow \text{TrapGen}(n, m, q)$.

2. For each attribute $i \in S^*$, $\mathcal{B}$ draws $(\mathbf{A}_i, \mathbf{v}_i) \in Z_q^{n\times m} \times Z_q^n$ from $O$, and set $\mathbf{A}_i = \mathbf{A}_i - \mathbf{B}$ in the public parameter;

3. For each attribute $i \notin S^*$, challenger $\mathcal{B}$ draws $(\mathbf{A}_i, \mathbf{v}_i) \in Z_q^{n\times m} \times Z_q^n$ from $O$, and set $\mathbf{A}_i = \mathbf{A}_i$ in the public parameter;

Finally, challenger $\mathcal{B}$ set the public parameter $pp = (\mathbf{A}_0, \mathbf{B}, \{\mathbf{A}_i\}_{i \in U}, \mathbf{s})$ and keep $msk = (\mathbf{T_B}, v_s, \mathbf{v}_0, \{\mathbf{v}_i\}_{i \in U})$ as secret.

**Key generation Queries1.** Now the challenger $\mathcal{B}$ has to produce a private key about the access policy $(M, \rho)$, but for this access policy $(M, \rho)$, the target attribute set $S^* = \{\text{Attrib}_1, \text{Attrib}_2, \cdots \text{Attrib}_t\} \subseteq U$ is a not authorized set. Here we describe the process that the challenger $\mathcal{B}$ how to construct a private key about $(M, \rho)$ as follow (where $M \in Z_q^{l \times n}, \rho : [l] \to Z_q$):

1. Since $S^*$ is not authorized for $(M, \rho)$, there exists an attribute $\rho(\tau) \notin S^*$ for all $\tau \in [l]$ at least. Without loss of generality, we suppose $\rho(1) \notin S^*$. By setup algorithm know, the challenger B know the trapdoor $\mathbf{T_B}$ of $\mathbf{B}$. Then the challenger $\mathcal{B}$ can calculate the trapdoor $\mathbf{T}_{\mathbf{E}_{\rho(1)}}$ of $\mathbf{E}_{\rho(1)} = \mathbf{A}_0 \| (\mathbf{A}_{\rho(1)} + \mathbf{B})$ through algorithm $ExtBasis$.

2. The according the algorithm $ExtBasis(\mathbf{T}_{\mathbf{E}_{\rho(1)}}, \mathbf{E} = \mathbf{E}_{\rho(1)} \| \mathbf{E}_{\rho(2)} \| \cdots \| \mathbf{E}_{\rho(l)})$ output the trapdoor $\mathbf{T_E}$ of $\mathbf{E}$. Then the challenger $\mathcal{B}$ use trapdoor $\mathbf{T_E}$ to get private key $sk_{\rho(k)}$ of attribute $\rho(k), k \in [l]$ through calling the algorithm $\text{SamplePre}(\mathbf{E}, \mathbf{T_E}, \mathbf{s}_j, \sigma)$ (where $\mathbf{s}_j = \{\underbrace{0, 0 \cdots, 0}_{j-1}, s_j, \underbrace{0, \cdots 0}_{n-j}\}$). Namely

$$\mathbf{E} \cdot sk_{\rho(k)} = \begin{bmatrix} 0 \\ \vdots \\ s_k \\ \vdots \\ 0 \end{bmatrix}_{n \times 1}$$

**Challenge.** When adversary $\mathcal{A}$ submits $msg_0, msg_1 \in \{0,1\}$, challenger $\mathcal{B}$ randomly chooses $b \in \{0,1\}$, and computes the ciphertext about attributes set $S^*$:

$$C_0 = v_0 + \left\lfloor \frac{q}{2} \right\rfloor msg_b \in Z_q$$

$$C' = \mathbf{v}_0 \in Z_q^m$$

$$C_i^* = \mathbf{v}_i \in Z_q^m, i \in S^*$$

And return ciphertext $C^* = \left(C_0, \mathbf{C}', \{\mathbf{C}_i^*\}_{i \in S^*}\right)$

**Key generation Queries2.** $\mathcal{B}$ answers queries of $\mathcal{A}$ the same way it does in key generation queries1 with the only restriction that the target attributes $S^*$ is not authorized for $(M, \rho)$ queries.

**Guess.** After the query 2, adversary $\mathcal{A}$ output a bit $b'$ as a guess for $b$. If $b = b'$, $\mathcal{B}$ outputs 1, else outputs 0.

## 4.3 Parameter

Because our scheme refer some classical algorithms and consider the correct and security of our scheme, we want the set the parameter appropriately.

1. Use TrapGen (see section 2.3 TrapGen) in setup, and the algorithm need $m \geq 5n \log q$;

2. Use SamplLefte (See section 2.5.5) in key generation, and need Gaussian parameter $\sigma > \|\tilde{\mathbf{T}}_{\mathbf{A}_0}\| \cdot \omega\left(\sqrt{\log 2m}\right)$;

3. For the hardness of LWE (see Proposition 8) assumption, and need $\alpha q \geq 2\sqrt{m}$;

4. For the SamplePre (see section 2.5.2) used in key generation queries of security proof, and need $\sigma \geq \|\tilde{\mathbf{T}}_{\mathbf{E}}\| \cdot \omega\left(\sqrt{\log lm}\right)$.

For the correctness of our scheme we need $\left| \chi_0 - \sum_{j \in I} \left( \sum_{i=1}^{l} \omega_i(\boldsymbol{\chi}) \mathbf{e}_i^{(j)} \right) \right| < \frac{q}{5}$. Combine the restriction about parameter above, we set the parameter in our scheme as follows:

Note that according to the TrapGen algorithm we know $\|\tilde{\mathbf{T}}_{\mathbf{A}_0}\|, \|\tilde{\mathbf{T}}_{\mathbf{E}}\| \leq O\left(\sqrt{n \log q}\right)$, and then we calculate $\left| \chi_0 - \sum_{j \in I} \left( \sum_{i=1}^{l} \omega_i(\boldsymbol{\chi}) \mathbf{e}_i^{(j)} \right) \right|$.

$$\left| \chi_0 - \sum_{j \in I} \left( \sum_{i=1}^{l} \omega_i(\chi) \mathbf{e}_i^{(j)} \right) \right|$$

$$\leq |\chi_0| + \left| \sum_{j \in I} \left( \sum_{i=1}^{l} \omega_i(\chi) \mathbf{e}_i^{(j)} \right) \right|$$

$$\leq |\chi_0| + \sum_{j=1}^{l} \left| \left( \sum_{i=1}^{l} \omega_i(\chi) \mathbf{e}_i^{(j)} \right) \right|$$

$$\leq |\chi_0| + l^2 \left| \chi \mathbf{e}_i^{(j)} \right|$$

$$\leq q\alpha\omega\left(\sqrt{\log 2m}\right) + \frac{1}{2} + l^2 \left( \left\| \mathbf{e}_i^{(j)} \right\| q\alpha\omega\left(\sqrt{\log 2m}\right) + \left\| \mathbf{e}_i^{(j)} \right\| \frac{\sqrt{2m}}{2} \right)$$

$$\leq \left( q\alpha\omega\left(\sqrt{\log 2m}\right) + \frac{1}{2} \right) + l^2 \sigma \sqrt{2m} \left( q\alpha\omega\left(\sqrt{\log 2m}\right) + \frac{\sqrt{2m}}{2} \right)$$

$$\leq \left( q\alpha\omega\left(\sqrt{\log 2m}\right) + \frac{\sqrt{2m}}{2} \right)\left(1 + l^2 \sigma \sqrt{2m}\right)$$

$$\leq \left( q\alpha\omega\left(\sqrt{\log 2m}\right) + \frac{q\alpha}{4} \right)\left(1 + l^2 \sigma \sqrt{2m}\right)$$

$$= q\alpha\left( \omega\left(\sqrt{\log 2m}\right) + 1 \right)\left(1 + l^2 \sigma \sqrt{2m}\right)$$

So we set the parameter such that $q\alpha\left(\omega\left(\sqrt{\log 2m}\right)+1\right)\left(1+l^2\sigma\sqrt{2m}\right) \leq \frac{q}{5}$ and satisfy four restrictions above, our scheme will be success with overwhelming probability.

Assume that $\delta$ is a real such that $n^{1+\delta} > n\log q$, and set $n, m, \sigma, q, \alpha$ as follow:

1. Take $n$ such that $n^{1+\delta} > n\log q$ appropriate;

2. Set $m = n^{1.5} \geq 5n\log q$, satisfying 1 above;

3. $\sigma = lm \cdot \omega(\log lm)$, satisfying 2 and 4 above;

4. The noise parameter $\alpha = \frac{1}{5}\left(\left(\omega\left(\sqrt{\log 2m}\right)+1\right)\left(1+l^2\sigma\sqrt{2m}\right)\right)^{-1}$ satisfying 3 above;

5. Set large prime $q$ as $q \geq 10\sqrt{2m}\left(\omega\left(\sqrt{\log 2m}\right)+1\right)\left(1+l^2\sigma\sqrt{2m}\right)$, satisfying the inequation.

For this setting, it is not only satisfying the condition algorithm we used but also can decrypt ciphertext correctly with overwhelming probability.

## 5. Conclusions

In this paper, we present a large universe attribute-based encryption from lattices, and the size of the attribute sets can be exponentially large, and we can add attributes after the system setup. The scheme only uses a matrix as the master private key, it makes the master private key's storage cost is too small; moreover, the ciphertext divide into three parts in the encryption process, and to do so both reduce the size of the ciphertext and shorten the time encryption; Finally, under LWE assumption, our scheme is security under the select attribute attack.

**Acknowledgements**

This work is supported by the National Natural Science Foundation of China under grants 61173192 and 61303223, Research Foundation of Education Department of Shaanxi Province of China under grants 12JK0740. Thanks also go to the anonymous reviewers for their useful comments.